\documentclass[12pt]{iopart}

\expandafter\let\csname equation*\endcsname\relax
\expandafter\let\csname endequation*\endcsname\relax
\usepackage{amsmath,alltt}
\usepackage[psamsfonts]{amssymb}
\usepackage[dvips]{graphicx}
\usepackage{multirow}
\usepackage{rotating}
\usepackage{lscape}

\begin{document}

\title[A novel mechanism for the distance-redshift relation]{A novel mechanism for the distance-redshift relation}
\author{Nathan W. C. Leigh$^{1}$ and Or Graur$^{1,2,3}$} 

\address{$^{1}$Department of Astrophysics, American Museum of Natural History, Central Park West and 79th Street, New York, NY 10024}
\address{$^{2}$Harvard-Smithsonian Center for Astrophysics, 60 Garden St., Cambridge, MA 02138}
\address{$^{3}$NSF Astronomy \& Astrophysics Postdoctoral Fellow}
\ead{nleigh@amnh.org}

%




\begin{abstract}

\noindent We consider a novel mechanism to account for the observed distance-redshift relation.  This is done by presenting a toy model for the large-scale matter distribution in a static Universe.  Our model mainly concerns particles with masses far below those in the Standard Model of Particle Physics.  The model is founded on three main assumptions:  (1) a mass spectrum dN$_{\rm i}$/dm$_{\rm i}$ $=$ $\beta$m$_{\rm i}^{-\alpha}$ (where $\alpha$ and $\beta$ are both positive constants) for low-mass particles with m$_{\rm i}$ $\ll$ 10$^{-22}$ eV $\ll$ M$_{\rm P}$, where M$_{\rm P}$ is the Planck mass; (2) a particle mass-wavelength relation of the form $\lambda_{\rm i} =$ $\hbar$/$\delta_{\rm i}$m$_{\rm i}$c, where $\delta_{\rm i} =$ $\eta$m$_{\rm i}^{\gamma}$ and $\eta$ and $\gamma$ are both constants; and (3) For such low-mass particles, locality can only be defined on large spatial scales, comparable to or exceeding the particle wavelengths.

We use our model to derive the cosmological redshift characteristic of the Standard Model of Cosmology, which becomes a gravitational redshift in our model.  We compare the results of our model to empirical data and show that, in order to reproduce the sub-linear form of the observed distance-redshift relation, our model requires $\alpha$ + $\gamma$ $<$ 0.  
We further place our toy model in the context of the Friedmann Universe via a superposition of Einstein Universes, each with its own scale factor a$_{\rm i}$.  Given the overwhelming evidence supporting an expanding Universe, we then address possible modifications to our base model that would be required to account for the available empirical constraints, including the addition of some initial expansion.  
Finally, we consider potentially observable distinctions between the cosmological redshift and our proposed mechanism to account for the observed distance-redshift relation.
\end{abstract}

\maketitle


\section{Introduction} \label{intro}

\subsection{The distance-redshift relation}

The observed distance-redshift relation is a cornerstone of the current cosmological 
paradigm.  The matter distribution throughout the Universe is observed to be 
homogeneous and isotropic on large spatial scales.  Theoretically, the mean mass 
density decreases with increasing proper time due to the presence of the scale factor 
$a(t)$ in the Robertson-Walker metric.  It is the Robertson-Walker scale factor 
that drives the expansion of the Universe in the Standard Model of Cosmology, 
called $\Lambda$CDM \cite{weinberg08}.  This is manifested observationally 
in the form of a cosmological redshift, or Hubble's Law: distant galaxies at low redshift 
(z $\ll$ 1) appear to be receding with a recession velocity that is linearly proportional 
to their distance from us \cite{hubble29,riess09}. 

At large redshifts (z $\gtrsim$ 0.6), 
the observed distance-redshift relation begins 
to deviate significantly from linearity and becomes noticeably sub-linear \cite{amanullah10,hinshaw13,ade15}.\footnote{We 
define this limit as follows.  We fit a straight line to the data (in linear-linear, instead of linear-log, space), and 
calculate the corresponding reduced chi-squared.  We then begin to restrict the range of redshifts (by excluding data 
points with redshifts greater than a given upper limit) until the reduced chi-squared drops below unity.  The upper limit 
for the redshift corresponding to a reduced chi-squared of unity defines the point at which the distance-redshift 
relation starts to deviate from linearity.}  
This observed acceleration in the expansion of the Universe at the present 
epoch is attributed to a mysterious dark energy, whose nature is unknown (see \cite{frieman08} for 
a review).

The distance-redshift relation is interpreted as evidence for an expanding Universe.  Indeed, a hot and dense 
early state for the Universe has now been firmly established, and the evidence is 
extensive.  For example, the standard surface brightness tests have been performed \cite{lubin01}, which involves 
comparing the luminosity distance of a given source with its angular diameter distance.  Other evidence 
comes from observations of the x-ray luminosities of galaxy clusters, intergalactic absorption 
measurements, galaxy number counts, cosmic abundances, etc. (for more details see Chapter 1 
of \cite{weinberg08} and references therein).  Perhaps the most compelling evidence for a hot and dense 
early Universe comes from the cosmic microwave background, which has spurred a great deal of new 
cosmological data since its discovery in 1965 \cite{penzias65}.

In this paper, we present a novel mechanism to account for the observed distance-redshift relation.  
To this end, we introduce a toy model for the large-scale matter distribution in the Universe.  Our model 
considers the possibility that, at very low particle masses (well below any particle masses in 
the Standard Model of Particle Physics), locality can only be defined on large spatial scales.  
Extremely light (m $\sim$ 10$^{-22}$ eV) bosons have been proposed as 
dark matter candidates, with de Broglie wavelengths $\lambda \sim$ 1 kpc (see, for example, 
\cite{hui16} and references therein).  Often called "fuzzy dark matter", this alternative to cold 
dark matter could explain various observational discrepancies with the predictions of $\Lambda$CDM, 
such as the existence of globular clusters (GCs) in the Fornax dwarf galaxy.  These GCs should 
have spiraled in to the nucleus long ago due to dynamical friction in a cold dark matter (CDM) halo.  An intriguing 
extension of this idea could, for instance, be applied to an even lighter particle having similar 
effects in galaxy clusters and/or groups.  Importantly, the existence of such ultralight particles would 
have been largely over-looked as astrophysically significant, since any inter-particle interactions 
would occur at very low energies well outside the range of detectability.

As we will explain, our proposed mechanism for the observed distance-redshift relation operates 
without altering many of the successes of the current Standard Model for Cosmology in accounting 
for the available observational constraints, while also avoiding some of its theoretical 
uncertainties (e.g., the singularity at t $=$ 0).  

\subsection{Mass density}

The concept of mass density permeates a number of physical 
fields, and is at the forefront of some of the most challenging puzzles of modern 
astrophysics.  On 
large spatial scales, the issue of mass density is related to several cosmological 
paradigms, including both dark matter and dark energy.  
On very small scales, mass density is a theme central to the development of the 
unification of quantum mechanics (QM) and general relativity (GR), called quantum 
gravity \cite{burgess04,donoghue94}.  
A sticking point with quantum gravity theories is how to model the interaction between matter and 
space-time at spatial scales smaller than the Planck length.
Thus, advancing our 
understanding of mass density could be crucial to future progress in several 
sub-disciplines within both physics and astronomy.  

\begin{figure}
\begin{center}
\resizebox{!}{85mm}{\includegraphics{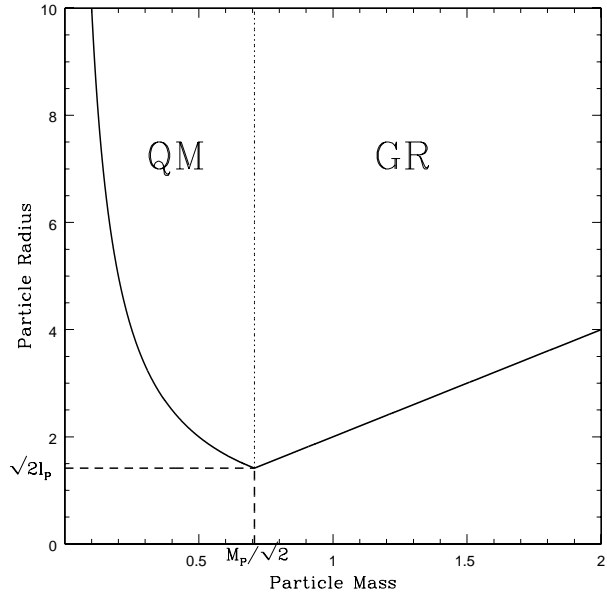}}
\caption{An upper limit for the effective particle radius, or characteristic wavelength $\lambda$, is shown 
as a function of the particle mass $m$, as given by Equation~\ref{eqn:psize1}.  
In making this figure, we have assumed $G = c = \hbar = 1$.
\label{fig:fig1}}
\end{center}
\end{figure}

Below the Planck scale, the uncertainty principle becomes important, and all known physical 
theories break down.  The classical example of this is shown 
in Figure~\ref{fig:fig1}, in 
which we adopt, for illustrative purposes, the Compton wavelength and the 
Schwarzschild radius as lower and upper limits, respectively, for the characteristic 
``particle'' wavelength or radius $\lambda$ below and above the Planck limit, 
respectively.  That is:\footnote{The transition mass 
m $=$ M$_{\rm P}$/$\sqrt{2}$ is found by setting R$_{\rm S} = \lambda_{\rm c}$, and solving for m.}
\begin{eqnarray}
\label{eqn:psize2}
\lambda &=& \frac{\hbar}{mc}, m \ge M_{\rm P}/\sqrt{2} \\
        &=& \frac{2Gm}{c^2}, m \le M_{\rm P}/\sqrt{2}
\end{eqnarray}  
Equation~\ref{eqn:psize2} marks the intersection between quantum mechanics and general relativity, 
and the point at which these two physical theories break down.  
In making Figure~\ref{fig:fig1}, we have assumed that Planck mass black holes (BHs) are stable,
and have thus ignored the emission of Hawking radiation \cite{hawking74}.  

This choice for $\lambda$ is motivated by the fact that the notions of elementary 
particle and BH are thought to merge below the Planck scale \cite{thooft85}.  
This is supported by the fact that the Compton wavelength $\lambda_{\rm c} = \hbar$/mc 
becomes on the order of the Schwarzschild radius R$_{\rm S} =$ 2Gm/c$^2$ at these 
small scales, and quantum fluctuations in the position of the black hole affect the very 
definition of the horizon \cite{coleman92}.  \textit{The key point to take away from Equation~\ref{eqn:psize2} 
is that, below the Planck scale, the particle wavelength is inversely proportional to its mass.}

\subsection{Gravitational collapse and singularity formation}

An arguably critical example of the limitations imposed by the concept of mass density is 
the formation of singularities, or objects of 
infinite mass density.  The nature of singularities, which represent a limiting density 
at which the metric tensor in the Einstein Field equations is 
undefined \cite{landau75}, is unknown.  That is, 
continuous differentiable manifolds predict infinite curvature at singular
points, indicating the breakdown of GR at very small spatial scales.

And yet, many authors have argued that true physical singularities 
do exist in nature.  For instance, it was first argued by 
\cite{oppenheimer39} that, for a pressure-free 
spherical distribution of matter, the final fate of gravitational 
collapse is a true physical singularity that cannot be removed by 
any coordinate transformation.  This result was generalized by 
\cite{penrose65}, who argued that the 
assumption of spherical symmetry is not needed to ensure 
that matter collapses to a singularity.  \cite{hawking76}, 
among others, later argued that the breakdown of the classical concepts 
of space and time associated with the formation of singularities represents a 
fundamental limitation in our ability to predict the future, in analogy with (but 
additional to) the limitations imposed by the uncertainty principle in 
QM.  However, 
causality need not break down if an event horizon prevents singularities 
from ever being observed by the external Universe.  Indeed, this seems to suggest 
that, with the exception of the Big Bang singularity in cosmology, no naked 
singularities should exist in nature \cite{penrose69}. 

The physical significance of the breakdown of GR at the Planck scale 
is not yet understood.  For example, in the case of the Robertson-Walker metric, 
there exist different sets of coordinates describing the manifold at the 
t$=0$ singularity.  Depending on the choice of coordinates, the 
singularity can be modeled either as a three-surface or a singular 
point \cite{weinberg08}.  More generally, different
manifold structures can be adopted to model singularities
that often agree for non-singular regions but disagree at the singular
points.  

\subsection{The observer in cosmological models: consistency over many orders of magnitude}

Any successful model for the large-scale structure of the Universe must be 
founded on assumptions that remain valid over many orders of magnitude in space and time.  
In cosmology, at early times, the assumptions underlying GR and the application of Einstein's 
equations must 
remain valid in the domain where quantum mechanical effects become non-negligible.  
These effects are generally thought to only be important on very small spatial scales 
\cite{donoghue94b}, corresponding to large matter densities.  Given enough mass, these 
small-scale quantum mechanical effects, acting like a repulsive force during gravitational collapse, 
can be overcome (e.g., a super-Chandrasekhar mass white dwarf and the inverse beta decay events that must 
occur to facilitate collapse to a neutron star).  More generally, at microscopic distance scales, quantum 
mechanics can lead to a modification of 
the gravitational potential.  But it is not always clear how to treat the quantum state of the matter 
sourcing the energy-momentum tensor T$_{\rm \mu\nu}$ in the Einstein equations. 

Importantly, the time-reversal of these transitions 
are neglected as directly affecting the observer in cosmological models.  
That is, the (mean) mass density of the Universe exceeds the particle mass density at very early times.  
If this ever truly occurs in nature, from Figure~\ref{fig:fig1}, it must be inside an event horizon.  
Hence, empirically, this does not occur in the visible Universe.  Said another way, 
Big Bang cosmology, and the existence of singularities in general, implies that at some 
point in the distant past the space-time containing any observer must have been part of the 
very system the observer is measuring.  It follows that a fully consistent quantum mechanical 
description of the very early Universe, and the role of the observer, should be applied.  
It is not completely clear how to properly accommodate these issues within the framework of 
cosmological models.  

The key point is that models should not be extended past the domain of their validity.  Beyond 
these critical points, crucial assumptions break down.  In cosmology, it is (arguably) the nature of the 
observer that must be able to properly accommodate the transitional points in space and time 
described above.  This should 
perhaps come as no surprise, given the many orders of magnitude in scale that must be crossed.  
Our motivation for re-examining in this paper the redshift-distance relation and the large-scale 
distribution of mass in the Universe originates from these issues with the current Standard Cosmological Model.  

In this paper, we consider a novel mechanism to account for the observed distance-redshift relation.  This is 
described by presenting a toy model for the large-scale matter distribution in a static 
(i.e., non-expanding) Universe.  Our model mainly concerns very low-mass particles with 
masses far below those of the Standard Model of Particle Physics (m $\le$ 10$^{-22}$ eV $\ll$ M$_{\rm P}$), 
since here the characteristic particle 
wavelengths could be comparable to the immense spatial scales of interest.  Hence, our model 
is effectively motivated by an extrapolation of known physical concepts to astrophysical scales.  By 
necessity, we make a number of assumptions in deriving our model, and address any speculative 
aspects of these assumptions via a discussion of their validity and implications for large-scale 
astrophysics.  Given these critical assumptions, we show in Section~\ref{model} that the 
cosmological redshift characteristic of the Standard Model of Cosmology becomes a 
gravitational redshift in our toy model.  We then use our model to derive Hubble's Law 
and highlight a few potentially observable 
distinctions between our model and the predictions of $\Lambda$CDM.  We further incorporate our 
toy model into the Friedmann Universe, in order to better understand the characteristic behavior expected for a 
more dynamic version of our model, as well as possible modifications to our base model needed to reproduce the 
available empirical constraints.  In Section~\ref{discussion}, we discuss the possible significance of our 
results for the observed distance-redshift relation and, more generally, cosmological models.  Our key conclusions 
are summarized in Section~\ref{summary}.

\section{Model} \label{model}

In this section, we present a new mechanism to account for the observed distance-redshift relation, via a toy 
model for the large-scale matter distribution in a static (i.e., non-expanding) 
Universe.  Using our model, we calculate the redshift of a photon emitted by a distant source and derive the 
predicted distance-redshift relation.  We begin with the assumption of a linear distance-redshift relation, 
in order to first reproduce Hubble's Law, but later we relax the assumption of linearity.  We go on to compare 
the predictions of our model to the observed distance-redshift relation, which we show constrains 
the distribution of particle masses in our model (i.e., the low-mass particle mass function).

For simplicity, throughout this section, we discuss our model mainly in the context of Euclidean space, and defer 
a discussion of relativistic effects to Section~\ref{friedmann} and Section~\ref{discussion}.

\subsection{Redshift} \label{redshift}

Consider an observer who wishes to measure the mass distribution of the Universe on large spatial 
scales.  We adopt a \textit{static} (i.e., non-expanding) toy model for the Universe, taken in the frame of reference of a particular 
particle (or wave packet) for simplicity.   Our particle has mass m$_{\rm 1} \le$ 10$^{-22}$ eV $\ll$ M$_{\rm P}$ and characteristic wavelength $\lambda_{\rm 1} \gtrsim$ 1 kpc $\gg$ l$_{\rm P}$ \cite{hui16} given by:  
\begin{equation}
\label{eqn:psize1}
\lambda_{\rm 1} = \frac{\hbar}{\delta_{\rm 1}m_{\rm 1}c},
\end{equation} 
where $\delta_{\rm 1} = \delta_{\rm 1}$(m$_{\rm 1}$) is a function of the particle mass satisfying 
0 $\le$ $\delta_{\rm 1}$ $\le$ 1.  

We adopt a continuous mass spectrum of 
particle masses m$_{\rm i}$, with m$_{\rm i+1}$ $<$ m$_{\rm i}$ and $\lambda_{\rm i+1} > \lambda_{\rm i}$ 
for all i.  That is:
\begin{equation}
\label{eqn:mf}
\frac{dN_{\rm i}}{dm_{\rm i}} = {\beta}m_{\rm i}^{-\alpha},
\end{equation}
where $\alpha$ and $\beta$ are both positive constants.  We also adopt the following functional form for the particle 
mass-wavelength relation:\footnote{Note that, if the particles are relativistic, the Lorentz factor $\gamma_{\rm i}$ should be 
included in the denominator of Equation~\ref{eqn:psize}.  However, for the time being, this can effectively be absorbed 
into $\delta_{\rm i}$, which is a free parameter in our model.  We will return to the implications of including relativistic 
effects in our model in Section~\ref{discussion}.} 
\begin{equation}
\label{eqn:psize}
\lambda_{\rm i} = \frac{\hbar}{\delta_{\rm i}m_{\rm i}c} > \lambda_{\rm c}
\end{equation}  
Here, $\delta_{\rm i} = \delta_{\rm i}$(m$_{\rm i}$) is 
a function of the particle mass satisfying 0 $\le$ $\delta_{\rm i}$ $\le$ 1, with $\delta_{\rm i} =$ 1 corresponding to 
the Compton wavelength $\lambda_{\rm c}$, which is in some cases a reasonable lower limit for the particle radius 
(see Section~\ref{intro}).  Note that $\delta_{\rm i} <$ 1 is certainly possible, for example this is the case for the 
semi-classical limit for the electron radius at the electroweak scale.  Note that the particle mass function 
in Equation~\ref{eqn:mf} mainly concerns very low-mass particles, with masses far below those covered by the 
Standard Model of Physics.  Here, the characteristic particle wavelengths could be comparable to the immense 
spatial scales of interest.  
For each particle mass (i.e., for every value of i), we assume a constant value for the corresponding 
mean mass density $\epsilon_{\rm i}$ in the Universe, and require that $\epsilon_{\rm i+1} > \epsilon_{\rm i}$.  

We make one more key assumption in our model.  This is a stipulation on Gauss' Law, which is used to calculate the gravitational 
field corresponding to a particular matter distribution.  Only particles both (1) with the maximum of their wave function 
located within the boundary and (2) a characteristic wavelength $\lambda_{\rm i}$ smaller than the size of the bounded region 
are included as contributing to the matter distribution.  Otherwise, the particles do not have a measurable 
gravitational effect (within the bounded region).  Specifically, the mass enclosed within a volume of radius r can be 
written:
\begin{equation}
\label{eqn:gauss}
M(r) = 4{\pi}\int_0^r \epsilon_{\rm i}(r')r'^2 dr' \sim \frac{4{\pi}}{3}\epsilon_{\rm i}r^3, \lambda_{\rm i} < r < \lambda_{\rm i+1}
\end{equation}
where the approximation follows from the assumption that $\epsilon_{\rm i} \gg \epsilon_{\rm 1}$ 
(i.e., $\epsilon_{\rm i}$(r'$=$r) $\gg$ $\epsilon_{\rm 1}$(r'$=$0)).  

For example, consider a typical Milky Way globular cluster (GC).  Observationally, 
these objects do not contain significant amounts of dark matter.  Within the context of our model, this is the 
case provided $\lambda_{\rm 1}$ $\ll$ r$_{\rm GC}$ $\ll$ $\lambda_{\rm 2}$, where r$_{\rm GC}$ is the 
typical size of a GC.  Particles of mass m$_{\rm 1}$ act as gravitating objects within such clusters and 
contribute to the total gravitational potential, but particles of mass m$_{\rm 2}$ do not.  

Given the above assumptions, we now consider an event in which our particle receives a photon emitted from a 
source located at a distance r from our particle, with $\lambda_{\rm i+1} >$ r $> \lambda_{\rm i} \gg \lambda_{\rm 1}$.  
Given our assumption regarding 
Gauss' Law, the photon is effectively emitted from a region of constant mass density $\epsilon_{\rm i}$, but is received 
by an observer (i.e., our particle) who perceives a Universe with a mean mass density $\epsilon_{\rm 1}$, and 
$\epsilon_{\rm i}$ $\gg$ $\epsilon_{\rm 1}$.  Hence, the photon is subject to a gravitational redshift:
\begin{equation}
\label{eqn:redshift}
z = \frac{\lambda_{\rm 1,i} - \lambda_{\rm 1,1}}{\lambda_{\rm 1,1}},
\end{equation}  
where $\lambda_{\rm 1,1}$ is the wavelength of the photon as measured \textit{locally} by an observer 
or particle of mass m$_{\rm 1}$, 
and $\lambda_{\rm 1,i}$ is the wavelength of the photon as measured by the receiving particle.  

The ratio $\lambda_{\rm 1,i}/\lambda_{\rm 1,1}$ can be derived as follows.  First, we assume that every mass species 
self-virializes within a Hubble time.  Hence, at the present epoch, we have for the total (mechanical) energy in particles of 
mass m$_{\rm i}$ (within a specified volume):
\begin{equation}
\label{eqn:energy}
E_{\rm i} = -T_{\rm i} = \frac{1}{2}V_{\rm i}
\end{equation}
where T$_{\rm i}$ and V$_{\rm i}$ are the kinetic and potential energy, respectively, of particles with 
mass m$_{\rm i}$.
Taking V$_{\rm i}$ $\sim$ GM$_{\rm i}$/$\lambda_{\rm i}$ $\sim$ G$\epsilon_{\rm i}\lambda_{\rm i}^2$, we have:
\begin{equation}
\label{eqn:energy4}
\frac{\lambda_{\rm 1,i}}{\lambda_{\rm 1,1}} = \frac{E_{\rm i}}{E_{\rm 1}} \sim \frac{\epsilon_{\rm i}\lambda_{\rm i}^2}{\epsilon_{\rm 1}\lambda_{\rm 1}^2} \sim \frac{m_{\rm i}^{1-\alpha}\lambda_{\rm, i}^2}{m_{\rm 1}^{1-\alpha}\lambda_{\rm 1}},
\end{equation}
where the last equality holds since we are considering a specified volume.  Thus, in our model, Equations~\ref{eqn:redshift} 
and~\ref{eqn:energy4} replace the cosmological 
redshift in $\Lambda$CDM, which is generated indirectly via the Robertson-Walker scale factor.  

\subsection{Hubble's Law} \label{hubble}

Next, we derive Hubble's Law within the context of our simple model.  First, from Section~\ref{redshift}, we have:
\begin{equation}
\label{eqn:redshift2}
z \sim \frac{m_{\rm i}^{1-\alpha}\lambda_{\rm i}^2}{m_{\rm 1}^{1-\alpha}\lambda_{\rm 1}} - 1 \sim \Big( \frac{\lambda_{\rm i}}{\lambda_{\rm 1}} \Big)^{(\alpha+2\gamma+1)/(\gamma+1)} - 1
\end{equation}  

Now, a photon is emitted from a source located at a distance r from our observer or particle (located 
at r $=$ 0), and $\lambda_{\rm i+1} >$ r $> \lambda_{\rm i}$.  Hence, plugging Equation~\ref{eqn:psize} into 
Equation~\ref{eqn:redshift2} and assuming:  
\begin{equation}
\label{eqn:delta}
\delta_{\rm i} = {\eta}m_{\rm i}^{\gamma}, 
\end{equation}
where $\gamma$ and $\eta$ are both constants, we have:
\begin{equation}
\label{eqn:redshift3}
z \sim \Big( \frac{{\eta}m_{\rm 1}^{\gamma+1}c}{\hbar} \Big)^{(\alpha+2\gamma+1)/(\gamma+1)}r^{(\alpha+2\gamma+1)/(\gamma+1)}
\end{equation}
where the substitution r $\sim$ $\lambda_{\rm i}$ was made in the last equality.  

Hubble's Law gives:
\begin{equation}
\label{eqn:hubble2}
cz = H_{\rm 0}r,
\end{equation}
where c is the speed of light and H$_{\rm 0}$ is Hubble's constant.  \textit{We emphasize that, in our model, H$_{\rm 0}$ 
is merely a dummy variable or constant, and H$_{\rm 0} \ne \dot{a}$/a since we are considering a static model for the purposes 
of deriving the distance-redshift relation.}  In order to reproduce 
Equation~\ref{eqn:hubble2} in our model, we require that ($\alpha$+2$\gamma$+1)/($\gamma$ + 1) $=$ 1 or:
\begin{equation}
\label{eqn:hubble3}
\alpha + \gamma = 0
\end{equation}
and
\begin{equation}
\label{eqn:H0}
H_{\rm 0} = \frac{{\eta}m_{\rm 1}^{1-\alpha}c^2}{\hbar}
\end{equation}

\begin{figure}
\begin{center}
\resizebox{!}{95mm}{\includegraphics{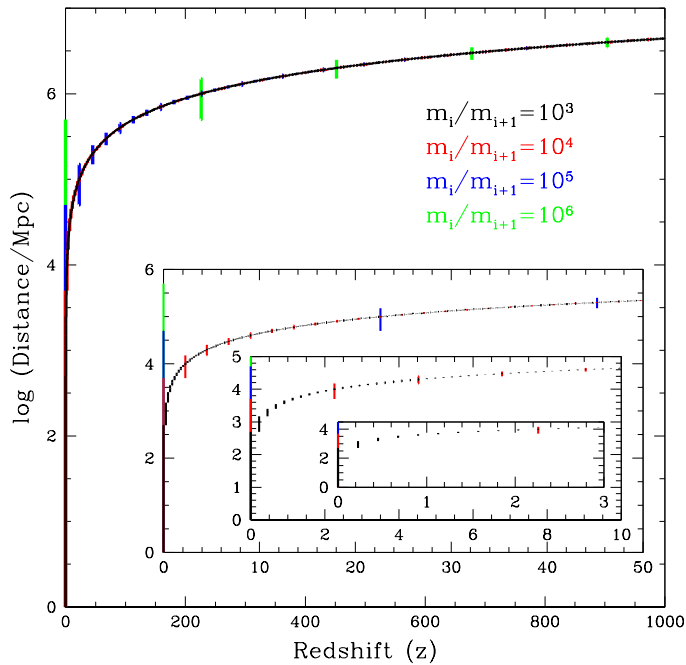}}
\caption{The distance-redshift relation predicted by our model, for different bin sizes in the 
particle mass function dN$_{\rm i}$/dm$_{\rm i}$, and assuming $\alpha =$ 1/2 and $\gamma =$ -1/2.  
Specifically, the black, red, blue and 
green lines correspond to constant bin sizes of m$_{\rm i}$/m$_{\rm i+1}$ $=$ 10$^3$, 10$^4$, 10$^5$ and 
10$^6$, respectively, for all i.  The horizontal dashes along the distance-redshift relations indicate 
the range of distances over which a given redshift should be observed.
\label{fig:fig2}}
\end{center}
\end{figure}

For illustrative purposes, we use our model to construct the distance-redshift relation shown in 
Figure~\ref{fig:fig2}, for different assumptions regarding the choice of bin size in the particle 
mass function 
dN$_{\rm i}$/dm$_{\rm i}$.  That is, we take $\alpha =$ 1/2 and $\gamma =$ -1/2, and we now 
assume a \textit{discrete} mass function with constant spacing between 
successive particle masses, or bin sizes, but vary the size of the bins.  Importantly, 
there is no known reason that the discretization of the particle mass function should assume 
a constant grid-spacing.  We make this assumption here for simplicity, but return to this 
important issue in Section~\ref{discussion}.  In making Figure~\ref{fig:fig2}, we adopt 
H$_{\rm 0}$ $=$ 67.8 $\pm$ 0.9 km/s/Mpc in Equation~\ref{eqn:redshift2} \cite{ade15}.

A few interesting features in Figure~\ref{fig:fig2} are worth noting.  First, our toy model predicts that 
only specific discrete redshifts 
should be observable in the distance-redshift relation, with the exact values depending on 
the details of the discretization of the particle mass function.  That is, for a given bin size or grid spacing, 
the colored horizontal lines in Figure~\ref{fig:fig2} mark where the observed data points should fall.  In 
the limit that the particle mass 
function is continuous, this discretization disappears and all redshifts are potentially observable.  
Second, our toy model predicts intrinsic dispersion in the observed distance-redshift relation, as 
shown by the horizontal lines in Figure~\ref{fig:fig2}.  At a given redshift, the magnitude 
of the dispersion should be proportional to the grid spacing in the particle mass function (i.e., the 
ratio m$_{\rm i}$/m$_{\rm i+1}$).  We emphasize that neither of these observed features in the 
distance-redshift relation are consistent with the predictions of $\Lambda$CDM cosmology.

\subsection{The observed distance-redshift relation} \label{observed}

In this section, we compare the predictions of our model to the observed distance-redshift relation.  
We assume Euclidean space for all our distance calculations.

\begin{figure}
\begin{center}
\resizebox{!}{95mm}{\includegraphics{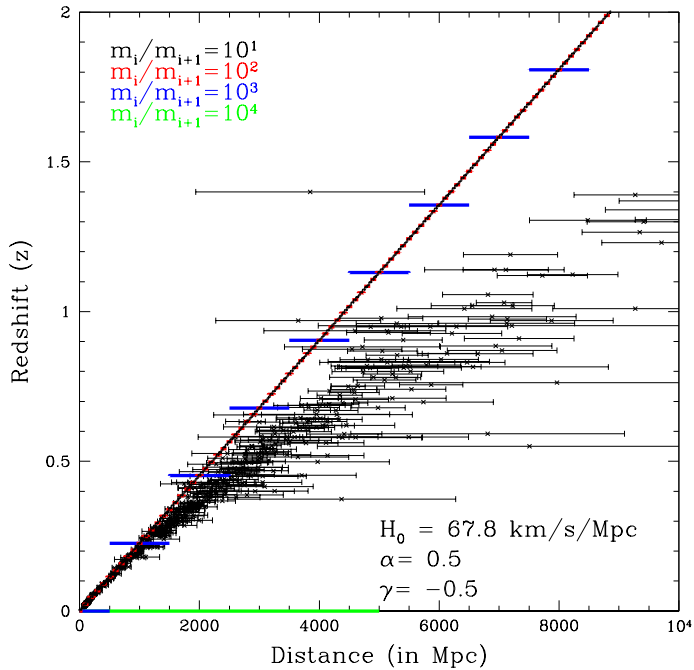}}
\caption{The distance-redshift relation predicted by our model, for different bin sizes in the 
particle mass function dN$_{\rm i}$/dm$_{\rm i}$, and assuming H$_{\rm 0}$ $=$ 67.8 km/s/Mpc,  
$\alpha =$ 1/2 and $\gamma =$ -1/2 in Equation~\ref{eqn:redshift2}.  Specifically, the black, red, blue and 
green lines correspond to constant bin sizes of m$_{\rm i}$/m$_{\rm i+1}$ $=$ 10$^1$, 10$^2$, 10$^3$ and 
10$^4$, respectively, for all i.  For comparison, we also plot with black crosses the observed data taken 
from the Union2 SN Ia compilation \cite{amanullah10}.  
\label{fig:fig3}}
\end{center}
\end{figure}

The discretization of the particle mass function is critical to predicting the observed appearance of 
the distance-redshift relation using our model.  This can be quantified empirically by looking 
for gaps in the measured values of redshift, along with intrinsic dispersion at a given redshift.  For 
example, in Figure~\ref{fig:fig3} we re-plot the distance-redshift relation obtained in our model and 
shown in Figure~\ref{fig:fig2}, but over a smaller range in redshift.  For comparison, we also plot observed 
data taken from the Union2 SN Ia compilation \cite{amanullah10}, which is compiled from 17 
different datasets.  All SNe were fit using the same light curve fitter and analyzed uniformly.

\begin{figure}
\begin{center}
\resizebox{!}{95mm}{\includegraphics{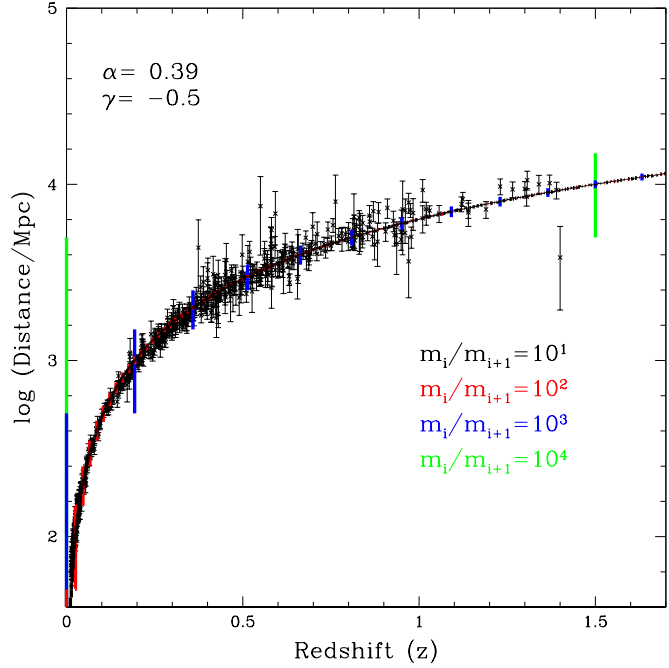}}
\caption{The same as in Figure~\ref{fig:fig3}, but adopting instead 
$\alpha =$ 0.39 and $\gamma =$ -0.5 in Equation~\ref{eqn:redshift2}.  Note that we plot redshift on the x-axis, and the 
logarithm of distance on the y-axis, since this is standard practice in the literature \cite{amanullah10}.
\label{fig:fig4}}
\end{center}
\end{figure}

A few things are apparent from a quick glance at Figure~\ref{fig:fig3}.  First, the observed 
distance-redshift relation is not linear; it appears to be slightly sub-linear.  Within 
the context of our model, this suggests that 
the quantity $(\alpha$ + 2$\gamma$ +1)/(1+ $\gamma$) should be slightly less than unity, or $\alpha$ + $\gamma$ $<$ 0.  
As illustrated in Figure~\ref{fig:fig4}, 
relaxing the assumption of a linear distance-redshift relation does indeed improve the agreement 
between our model and the observed data.  Figure~\ref{fig:fig4} shows that 
the data can be reasonably well matched by our model assuming 
$\alpha =$ 0.39 and $\gamma =$ -0.5.

Second, there does 
indeed appear to be intrinsic dispersion in the observed distance-redshift relation, but it is not 
clear whether or not this is due to observational uncertainties (not provided for all data points 
shown in Figure~\ref{fig:fig3}) or local gravitational effects.  Third, if taken at face value, these data 
suggest that there are no large gaps in the particle mass function and, very roughly, 
m$_{\rm i+1}$/m$_{\rm i}$ $<$ 10$^2$ for all i.  

We caution that our toy model could be too simple in its present form for direct comparisons to 
empirical data.  For 
instance, there is no reason to expect a constant binning in the particle mass function.  
We re-iterate here that 
the particle mass function is completely unconstrained, and other functional forms might also 
reproduce the observed distance-redshift relation in our model (such as, for example, a two- or 
three-part power-law).  What's more, we assume $\delta_{\rm i} =$ $\eta$m$_{\rm i}^{\gamma}$ for all i 
in Equation~\ref{eqn:psize} throughout this paper for simplicity, but remind the reader that 
this assumption is somewhat arbitrary.  Other assumptions for the value of $\delta_{\rm i}$ should 
directly affect the appearance of the distance-redshift relation predicted by our model.  

\section{Constructing a dynamic model} \label{dynamic}

In this section, we consider a more dynamic version of our simple toy model.  This will ultimately help us to 
assess modifications to our 
base model, which corresponds to a static Universe, needed to account for the overwhelming empirical 
evidence in favor of a hot and dense early state that quickly expanded, cooled and began re-condensing to 
form the observed large-scale structure of the present-day Universe. 

\subsection{The Friedmann Universe} \label{friedmann}

In this section, we place our toy model in the general framework of the Friedmann Universe.  This 
serves to further constrain the free parameters in our model, while also exploring the global implications of 
our model for the evolution of the underlying metric.  

The Cosmological Principle states that the metric for the Universe must take the general form:
\begin{equation}
\label{eqn:RWmetric}
ds^2 = a(ct)^2dl^2 - c^2dt^2,
\end{equation}
where dl$^2$ is a three-dimensional metric with constant curvature and a(ct) is the scale factor.  
Equation~\ref{eqn:RWmetric}, called the Robertson-Walker metric, can be plugged into Einstein's field 
equations, or:
\begin{equation}
\label{eqn:field}
R_{\rm \mu\nu} - \frac{1}{2}Rg_{\rm \mu\nu} = \frac{8{\pi}G}{c^4}T_{\rm \mu\nu},
\end{equation}
where T$_{\rm \mu\nu}$ is the energy-momentum tensor of the matter in the Universe, and must take 
the form of a perfect fluid in Robertson-Walker metrics.  This gives the Friedmann equations:
\begin{equation}
\begin{gathered}
\label{eqn:friedmann2}
\frac{2\ddot{a}}{a} + \frac{\dot{a}^2 + K}{a^2} = -\frac{8{\pi}G}{c^4}p \\
\frac{3(\dot{a}^2 + K)}{a^2} = \frac{8{\pi}G}{c^2}\bar{\epsilon},
\end{gathered}
\end{equation}
where p and $\bar{\epsilon}$ are the matter pressure and density, respectively, and K $=$ +1, -1, or 0 corresponds to the sign 
of the curvature. 

For a static cosmology, all time derivates in Equation~\ref{eqn:friedmann2} are zero.  This is the case in a (static) Einstein 
Universe.  In order to reproduce the observational constraint imposed by the data available to him at the time, Einstein 
introduced a cosmological constant $\Lambda$ into his model.  Here, in addition to the contribution from the gravitating 
matter (i.e., dust), the energy-momentum tensor contains a contribution proportional to the metric tensor:
\begin{equation}
\label{eqn:tensor}
\frac{8{\pi}G}{c^4}T_{\rm \mu\nu} = -{\Lambda}g_{\rm \mu\nu} + {\epsilon}u_{\rm \mu}u_{\rm \nu}, 
\end{equation}
where $\epsilon > 0$ and $\Lambda$ is a constant.  Using the relations:
\begin{equation}
\begin{gathered}
\label{eqn:friedmann3} 
\frac{8{\pi}G}{c^4}p = -\Lambda \\
\frac{8{\pi}G}{c^2}\bar{\epsilon} = \frac{8{\pi}G}{c^2}{\epsilon} + \Lambda,
\end{gathered}
\end{equation}
we obtain:
\begin{equation}
\begin{gathered}
\label{eqn:friedmann4} 
K = +1 \\
\Lambda = \frac{1}{a^2} \\
\frac{4{\pi}G}{c^2}\epsilon = \frac{1}{a^2},
\end{gathered}
\end{equation}
for an Einstein Universe.  Thus, the Einstein Universe is closed with constant curvature.

Now, in order to place the above in the context of our static toy model, consider the following.  First, we re-write 
Equation~\ref{eqn:RWmetric} in the form:
\begin{equation}
\label{eqn:RWmetric2}
ds^2 = (a_{\rm 0} - a_{\rm i})^2dl^2 - c^2dt^2,
\end{equation}
where a$_{\rm i}$ is the (constant) scale factor for particles of mass m$_{\rm i}$ and wavelength $\lambda_{\rm i}$, as 
given by Equation~\ref{eqn:psize}, and a$_{\rm 0}$ is a constant satisfying a$_{\rm 0} \ge$ a$_{\rm i}$ for all i.  Note that 
$\lambda_{\rm i} \le$ (a$_{\rm 0}$ - a$_{\rm i}$) for all i, with $\lambda_{\rm 1} \ll$ (a$_{\rm 0}$ - a$_{\rm 1}$) and 
$\lambda_{\rm i} \rightarrow$ (a$_{\rm 0}$ - a$_{\rm i}$) in the limit of very large i.  As we will show below, 
this parameterization is needed to ensure that the parameter $\alpha$ is positive.  Recall that, in our 
toy model, these particles observe a mean mass density $\epsilon_{\rm i}$ for the Universe, and 
a$_{\rm i+1} >$ a$_{\rm i}$ for all $i$.  For a pressureless dust (for example), the corresponding solutions to the Friedmann 
equations are then:
\begin{equation}
\begin{gathered}
\label{eqn:friedmann4} 
\Lambda_{\rm i} = \frac{1}{(a_{\rm 0} - a_{\rm i})^2} \\
\frac{4{\pi}G}{c^2}\epsilon_{\rm i} = \frac{1}{(a_{\rm 0} - a_{\rm i})^2},
\end{gathered}
\end{equation} 
and we assume a curvature of K $=$ +1 for every particle type i.  In a Friedmann Universe, the cosmological redshift 
is given by:
\begin{equation}
\label{eqn:redshift3}
z = \frac{\lambda_{\rm 2} - \lambda_{\rm 1}}{\lambda_{\rm 1}} = \frac{a(ct_{\rm 2})}{a(ct_{\rm 1})} - 1,
\end{equation}
for some times t$_{\rm 2} >$ t$_{\rm 1}$.  Hence, for our toy model, Equation~\ref{eqn:redshift3} becomes:
\begin{equation}
\label{eqn:redshift4}
z = \frac{a_{\rm 0} - a_{\rm 1}}{a_{\rm 0} - a_{\rm i}} - 1.
\end{equation}
Plugging Equation~\ref{eqn:friedmann4} into Equation~\ref{eqn:redshift4}, we obtain:
\begin{equation}
\label{eqn:redshift5}
z = \Big( \frac{\epsilon_{\rm i}}{\epsilon_{\rm 1}} \Big)^{1/2} - 1.
\end{equation}
A simple comparison with Equation~\ref{eqn:energy4} yields the constraint $\alpha =$ 1/2.  

It follows from this 
simple exercise that our toy model can be placed within the context of Friedmann's Universe via a superposition 
of Einstein space-times, each with its own scale factor a$_{\rm i+1} >$ a$_{\rm i}$.  As we will explain further 
below, using a quantity we call the particle packing fraction $F_{\rm p}$, a change 
in scale factor here can be interpreted as an increase in the ability of an observer (undergoing gravitational 
collapse) to resolve the spatial component of the line element.  From Equation~\ref{eqn:RWmetric2}, the unit of time 
in our chosen frame of reference is then determined by the wavelength crossing time, or the time taken by light 
to traverse one particle wavelength.  It is intriguing to consider whether or not our model might be able to use 
this "inverted" frame of reference, which corresponds to a particle (or collection of particles) belonging to a matter 
distribution crossing the high-energy barrier during gravitational collapse, to construct a cosmological 
model capable of (at least qualitatively) reproducing the relevant observations on large spatial scales.  

Cosmological perturbation theory can in principle be used to help justify our choice for the functional 
form of $\delta_{\rm i}$, as given by Equation~\ref{eqn:delta}.  For this, the perturbed geometry is often 
described in the general form:
\begin{equation}
\label{eqn:perturbed}
g_{\rm \mu\nu} = \bar{g}_{\rm \mu\nu} + {\delta}g_{\rm \mu\nu},
\end{equation}
where $\bar{g}_{\rm \mu\nu}$ is the unperturbed Friedmann metric and ${\delta}$g$_{\rm \mu\nu}$ corresponds to 
a small perturbation.  Through the Einstein equations, the metric perturbations should be coupled to perturbations in 
the matter distribution.  

Einstein's Universe is unstable to perturbations.  
Within the context of our model, however, any expansion will 
bring the particle wavelength $\lambda_{\rm i}$ into the space-time corresponding to the adjacent scale 
factor a$_{\rm i+1}$.  Here, the matter density $\epsilon_{\rm i+1} > \epsilon_{\rm i}$.  We speculate that 
this change in the balance between pressure and gravity should cause the expansion to 
reverse direction, and the perturbation should subsequently contract back into the space-time corresponding to its 
original scale factor a$_{\rm i}$.  Perturbative contraction, on the other hand, is free to proceed unimpeded.  This instability 
should ultimately 
allow for a cascading collapse scenario, where the matter distribution in the space-time corresponding to a given scale 
factor a$_{\rm i}$ collapses into the next a$_{\rm i+1}$, which collapses in to the next, and so on.  

\begin{figure}
\begin{center}
\resizebox{!}{95mm}{\includegraphics{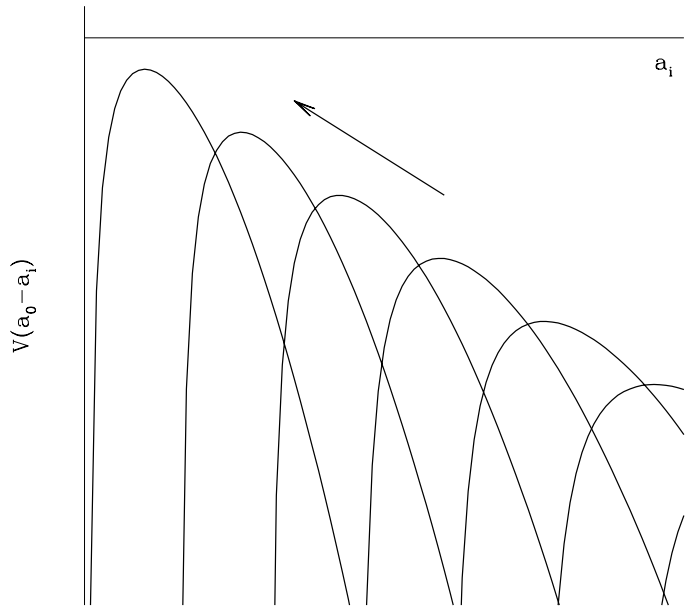}}
\caption{The potential energies V(a$_{\rm 0}$-a$_{\rm i}$) of our super-imposed Einstein Universes are shown as a function of their respective scale factors a$_{\rm i}$.  The arrow indicates the direction of gravitational collapse.
\label{fig:fig5}}
\end{center}
\end{figure}

The above very simple and qualitative picture is illustrated schematically in Figure~\ref{fig:fig5}, which shows the 
potential energies of our super-imposed Einstein Universes as a function of their respective scale factors a$_{\rm i}$.  By 
equating the potentials of adjacent scale factors at their points of intersection, or 
V(a$_{\rm i}$+${\delta}$a$_{\rm i}$) $=$ V(a$_{\rm i+1}$-${\delta}$a$_{\rm i+1}$), we obtain the trivial constraint:
\begin{equation}
\label{eqn:constraint1}
\epsilon_{\rm i} - {\delta}\epsilon_{\rm i} = \epsilon_{\rm i+1} + {\delta}\epsilon_{\rm i+1},
\end{equation}
provided a$_{\rm i}$+${\delta}$a$_{\rm i}$ $=$ a$_{\rm i+1}$-${\delta}$a$_{\rm i+1}$.

Although speculative, this dynamic version of our toy model shares a number of striking similarities to the observed 
large-scale structure of the Universe.  And yet, despite these successes, it seems likely that some further modifications 
of our base model will be needed to properly reproduce all available empirical constraints (e.g., baryon acoustic oscillations, 
the CMB, cosmic abundances, etc.).  For example, we would 
expect the collapse to occur where the particle wavelengths overlap in space 
and time.  This could ultimately give rise to a fractal structure resembling the large-scale cosmic web, but 
with a fractal dimension that is set by the ratio $\lambda_{\rm i}$/$\lambda_{\rm i+1}$, which is in turn set by the 
bin size between adjacent particle masses in the particle mass function.  Naively, the matter must be hot and dense 
when it condenses into the filaments in order to be able to reproduce, for example, the available constraints from 
cosmic abundance measurements.  Whether or not this simple picture could reproduce the observed CMB 
power spectrum and the effects of baryon acoustic oscillations as well is uncertain, and at least some initial 
expansion is likely needed in order to do so.  This expansion would occur 
each time matter condenses out of the space-time corresponding to a given scale factor a$_{\rm i}$ by 
collapsing gravitationally into the next a$_{\rm i+1}$; i.e., the matter from one scale factor a$_{\rm i}$ condenses out hot and 
dense before expanding into the space-time corresponding to the adjacent scale factor a$_{\rm i+1}$, cooling in the process 
before beginning to re-collapse gravitationally, and the process repeats.  This is illustrated in Figure~\ref{fig:fig5}.  

We emphasize that more work needs to be done to better understand the implications of cosmological 
perturbation theory for our model.  In particular, the metric given in Equation~\ref{eqn:RWmetric2} was chosen 
since it has the appropriate characteristic behavior to describe our model while also satisfying the Cosmological 
Principle.  Apart from this, the choice of metric is arbitrary and other metrics could also be considered.  We 
intend to explore these issues in future work, in an effort to better quantify further modifications to our base 
model that would be needed to properly reproduce the primary empirical constraints.

\subsection{Re-interpreting the scale factor} \label{scale factor}

In the Standard Model of Cosmology, 
the Robertson-Walker scale factor a(t) acts to decrease the mean mass density in the Universe with increasing proper time.  In our model, the Universe is static, and no 
expansion is needed at the present epoch to reproduce the observed distance-redshift relation.  
Hence, the volume of the observable Universe at t $=$ 0 is the same as at the present epoch (as 
observed by a particle of constant rest-mass), and 
an alternative mechanism is needed to decrease the mean mass density in 
the Universe with increasing proper time.  One way to do this is to adopt an appropriate frame 
of reference, specifically the frame of reference of a particular particle during gravitational collapse.  Here, 
changes in the Robertson-Walker scale factor are interpreted as an 
increase in the observer's ability to resolve the spatial component of the line element characteristic 
of the underlying space-time.  

Consider an initial state for the Universe at t $=$ 0 in which all particles have extremely low-masses, 
populating only the bottom-end of the particle mass function in Equation~\ref{eqn:mf}.  Gravity proceeds to 
dictate the time evolution of the Universe, causing particles to rapidly coalesce, merge and become more massive.  
This process continues unimpeded, quickly populating the full spectrum of particle masses in 
Equation~\ref{eqn:mf}.  Thus, our observing particle begins at t $=$ 0 with mass m$_{\rm i}$ and wavelength 
$\lambda_{\rm i}$, observing an initially dense Universe.  The particle eventually ends with mass 
m$_{\rm 1} \gg$ m$_{\rm i}$ and wavelength $\lambda_{\rm 1} \ll$ $\lambda_{\rm i}$, observing a much lower 
mean mass density in the Universe.  This occurs long before the present epoch, such that 
the seeds of structure formation are in place in the very early Universe.

\begin{figure}
\begin{center}
\resizebox{!}{95mm}{\includegraphics{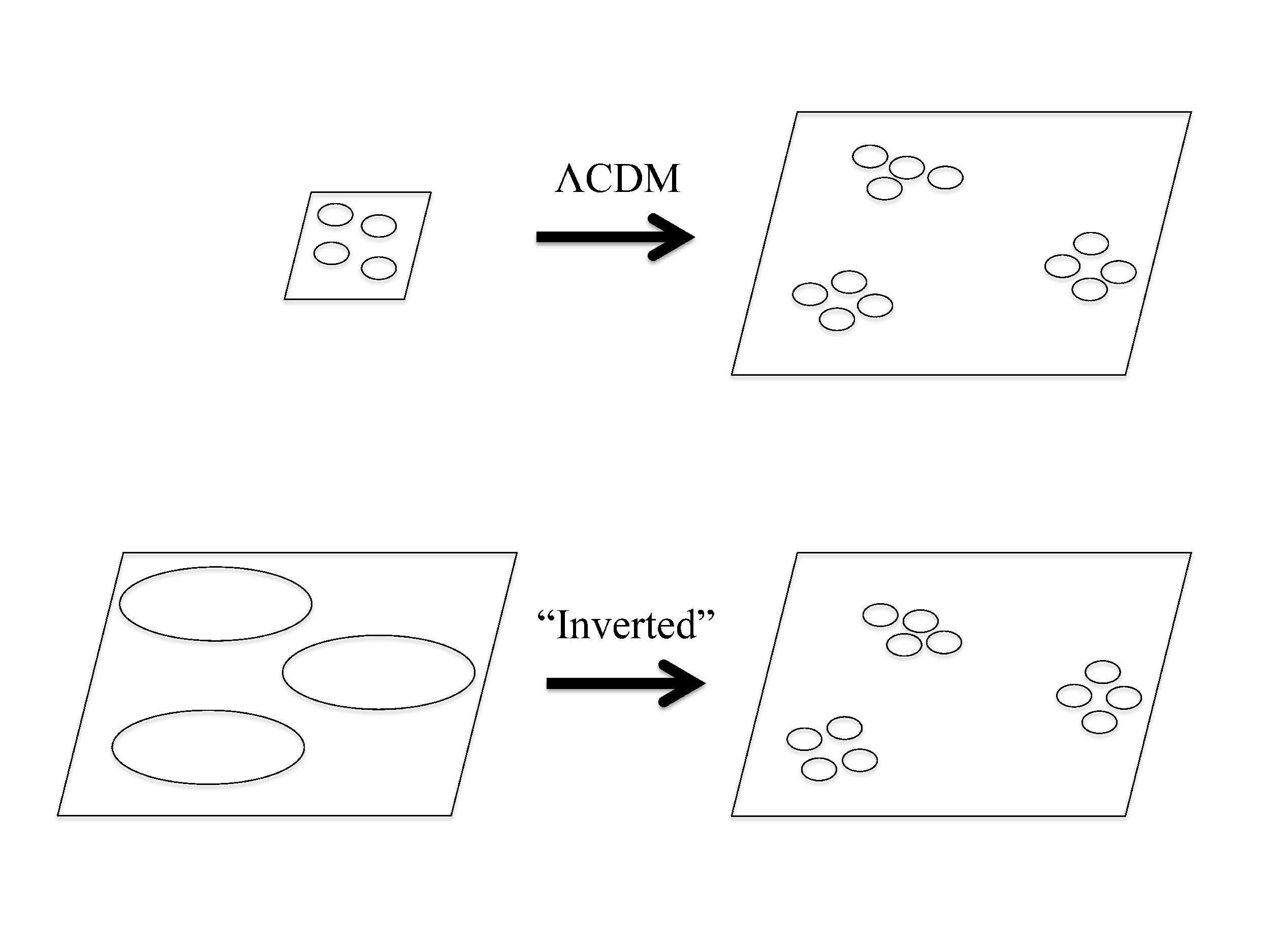}}
\caption{Schematic diagram representing the (zeroth-order) initial and final states in the Standard Model of Cosmology (top) 
and the "inverted" model discussed in the text (bottom).  In the Standard Model, space-time expands, leaving the particle rest-mass 
and wavelength unaffected.  In the "inverted" scenario, space-time is static, and particles begin with extremely low rest-masses 
and very long wavelengths, coalescing over time to form much more massive particles with much shorter wavelengths.  Note that 
the number of particles is not shown to scale; in particular, the number of particles should be highest in the bottom-left illustration.  
\label{fig:fig6}}
\end{center}
\end{figure}

In this scenario, illustrated schematically in Figure~\ref{fig:fig6}, there are two contributing factors 
to the perception of an expanding space-time or, equivalently, the perception of a mean mass density that decreases with increasing proper time.  First, by construction, particles can only exchange 
photons with other particles of the same mass, and $\epsilon_{\rm i+1} > \epsilon_{\rm i}$ for all i.  Hence, each 
time the particle rest-mass increases due to coalescence with other particles, the observing particle perceives a new 
particle distribution with a lower mean mass density.  Second, the perception of an expanding space-time could 
come from increasing the particle mass density directly in its own frame of reference (via direct particle-particle 
interactions), while holding the mean mass density of the Universe constant.  If 
the particle is unable to detect any change in its own mass density, then the result of this transformation in the 
particle reference frame is the perception of a decrease in the overall mean mass density of the Universe.  

To help illustrate this important point, consider the following parameter, 
which we call the particle packing fraction:\footnote{Classically, the packing fraction can be written 
F$_{\rm p}$ $=$ N$\Gamma_{\rm p}$/$\Gamma_{\rm 0}$ $=$ (Nm/$\Gamma_{\rm 0}$)$\Gamma_{\rm p}$/m 
$=$ $\epsilon_{\rm 0}$/$\epsilon_{\rm p}$, where N is the number of particles, m is the particle mass, $\Gamma_{\rm p}$ 
is the particle volume and $\Gamma_{\rm 0}$ is the volume of the container containing all N particles.}
\begin{equation}
\label{eqn:pack}
F_{\rm p,i} = \frac{\epsilon_{\rm i}}{\epsilon_{\rm p,i}},
\end{equation}
where $\epsilon_{\rm p,i}$ is the mean particle mass density (in the observing particle's own frame of reference) and 
$\epsilon_{\rm i}$ is the mean mass density of the Universe, as observed by particles with mass m$_{\rm i}$.  Importantly, 
the mean mass density $\epsilon_{\rm i}$ can only be \textit{indirectly} observed, by directly measuring the 
quantity F$_{\rm p,i}$.\footnote{In effect, in order for a measurement of a given quantity to hold any real meaning, a 
scale must first be defined by assigning units to the quantity or parameter in question.  Hence, in Equation~\ref{eqn:pack}, 
we are effectively measuring the mean mass density 
of the Universe $\epsilon_{\rm i}$ in units of the mean particle mass density $\epsilon_{\rm p,i}$.}  

In a Friedmann Universe, it is the Robertson-Walker 
scale factor a(t) that drives a decrease in $\epsilon_{\rm i}$ with increasing proper time, while the particle's 
own mass density $\epsilon_{\rm p,i}$ remains constant (see the top illustration in Figure~\ref{fig:fig6}).  
However, in the particle frame of reference, a decrease in 
F$_{\rm p,i}$ due to a decrease in $\epsilon_{\rm i}$ at constant $\epsilon_{\rm p,i}$ is 
equivalent to a decrease in F$_{\rm p,i}$ due to an increase in $\epsilon_{\rm p,i}$ at 
constant $\epsilon_{\rm i}$.  If the latter assumption is made, the time evolution of F$_{\rm p,i}$ 
must be driven by local changes in the particle mass density directly which 
must in turn be mediated by gravity.  Thus, in effect, the ``global expansion'' of 
space-time characteristic of the Standard Model of Cosmology is here replaced by a 
``local contraction.''  
That is, the quantity F$_{\rm p,i}$ decreases as the particle rest mass 
m$_{\rm i}$ increases, or as the observing particle ``slides down'' the particle mass function 
dN$_{\rm i}$/dm$_{\rm i}$.  Each time the particle's rest-mass increases via direct particle-particle interactions, 
it observes a new 
smaller mean mass density for the Universe $\epsilon_{\rm i}$.  We emphasize that this is analogous to the 
effect of increasing the Robertson-Walker scale factor with increasing proper time in $\Lambda$CDM cosmology.  
Thus, as shown in Section~\ref{friedmann}, in our toy model, increasing the Robertson-Walker 
scale factor can be interpreted as increasing the ability of the observer to resolve the spatial component of 
the line element of the underlying space-time.  

We emphasize that the above schematic or qualitative picture is far from a complete dynamic model, and 
relies on a number of idealized simplifying assumptions.  Nevertheless, this choice of reference frame is needed 
in our model to self-consistently bridge the orders upon orders of magnitude in space and time characteristic of 
the observable Universe.

\section{Discussion} \label{discussion}

In this section, we discuss the implications of our model for cosmology.  After briefly addressing some of 
the possible caveats and the expected impact of including additional relativistic effects in our model, we discuss 
potentially observable distinctions between the cosmological redshift and our proposed mechanism to account for the observed distance-redshift relation.  

\subsection{Relativistic effects} \label{relativistic}

First, we comment on the possible implications of including special relativistic corrections in our model, but 
emphasize that the magnitude of this effect is uncertain since the distributions of particle velocities are 
unknown.  If all particle species are assumed to be in energy equipartition in our model then, for extreme particle 
mass ratios, from this assumption it follows that the root-mean-square velocities of some 
very low-mass particles could become relativistic.  This is important since, in a model that includes relativistic 
effects, an additional Lorentz factor $\gamma_{\rm i}$ (where $\gamma_{\rm i} =$ 1/$\sqrt{1-\sigma_{\rm i}^2/c^2}$) 
must be included in the denominator of 
Equation~\ref{eqn:psize}.  Thus, large Lorentz factors contribute to a significant reduction in the particle 
wavelength, such that some fine-tuning would likely be required via the parameter $\delta_{\rm i}$, which is a 
free parameter that can be arbitrarily small in our model, in order to reproduce the observed data.  
Importantly, however, if the assumption of energy equipartition is relaxed, then the root-mean-square 
particle velocities need not be relativistic.  The overall qualitative results of our model are also independent 
of this assumption, which serves only to decrease the power-law index 
$\alpha$ in Equation~\ref{eqn:mf} by unity.  

More importantly, there is no known reason to expect energy 
equipartition in our model.  For example, an initial phase of gravitational collapse in the early 
Universe could be accompanied by violent relaxation, leaving the system out of thermal equilibrium.  
Whether or not the matter distribution in our model would have sufficient time to re-achieve energy 
equipartition is not clear.  Moreover, only the baryonic matter must ultimately pass through a hot and dense 
state in order to achieve consistency with the available primordial abundance constraints \cite{weinberg08}.  Any 
very low-mass particles contributing to large-scale gravitational potentials at the present-day (in our model)  
could not have passed through such a hot and dense state without relativistic effects having drastically reduced 
their characteristic wavelengths.  Nevertheless, the issue of the particle velocities (and hence wavelengths) is 
central to our toy model, which requires long wavelengths at very low particle masses in order to reproduce the 
available observational data.  This is an active area of research (see, for example, \cite{marsh15} and \cite{hui16}).


As for further adapting our model to include general relativistic effects, it is (in general) unclear how to source the energy-momentum 
tensor in the Einstein equations, since (among other things) the quantum state of the matter is unknown.  Finally, as already discussed, the discrete nature of our model could be difficult, if not impossible, to completely accommodate via Einstein's equations, since they are formulated from continuous and differentiable functions.  

\subsection{Empirical constraints} \label{empirical}

As explained in the preceding sections, our toy model bears many interesting similarities to an 
``inverted'' $\Lambda$CDM cosmology.  
But, as illustrated 
in Figure~\ref{fig:fig2}, several possible differences are also apparent.  In this section, we discuss 
potentially observable features of our model, and how they relate to both the available empirical data 
and theoretical models.

\subsubsection{Distance-redshift relation and Dark Energy} \label{dr}

$\Lambda$CDM predicts 
that (ignoring data uncertainties and local gravitational effects) all the data should fall precisely on 
the observed distance-redshift relation, with zero dispersion.  Conversely, in our toy model, 
we expect some intrinsic dispersion in the observed distance-redshift relation, with the magnitude 
of the dispersion being proportional to the grid spacing in the particle mass function (i.e., the ratio 
m$_{\rm i}$/m$_{\rm i+1}$; see Figure~\ref{fig:fig2}).  Next, our toy model predicts that only specific 
discrete values of the redshift should be observed in the distance-redshift relation, with the exact 
values depending on the details of the discretization of the particle mass function.  In the limit that 
the particle mass function is continuous, however, this potentially observable consequence of our 
model vanishes.  Importantly, the first observable feature (i.e. dispersion) is likely to offer a more practical constraint 
on our model.  This is because it would be difficult to establish that any gap detected in the 
distance-redshift relation is anything more than an observational bias, or selection effect.  Intrinsic 
dispersion, on the other hand, could be looked for by first finding a best-fit model for the data, adding (in 
quadrature) an intrinsic dispersion term $\sigma_{\rm int}$ to the uncertainties and 
calculating a reduced $\chi^2$ value.  If the reduced $\chi^2$ is less than or equal to the 
number of degrees of freedom in the model assuming $\sigma_{\rm int} =$ 0, then the data are 
consistent with having zero intrinsic dispersion.  If, on the other hand, we require $\sigma_{\rm int} >$ 0 
for an acceptable reduced $\chi^2$, then this could be used to constrain the degree of intrinsic 
dispersion in the data and, consequently, the bin size (i.e., the ratio of successive particle masses in the 
particle mass function) for the particle mass function.  We have attempted this simple test and 
find that, over the entire observed range of redshifts, the data are consistent with zero intrinsic 
dispersion.  However, this is not particularly telling, since we might only expect intrinsic dispersion 
to appear over a very narrow range of redshifts.  We conclude that a more sophisticated 
statistical treatment based around this method but confined to narrow ranges in redshift will be required 
to properly address this issue.

We have shown that the simple toy model presented here can potentially reproduce the 
observed shape of the distance-redshift relation at z $>$ 0.6 \cite{riess04}, presently attributed to 
dark energy in the Standard Model of Cosmology.  
However, our results suggest that significant fine-tuning 
is likely required via the parameter $\delta_{\rm i}$ in Equation~\ref{eqn:psize} in order to avoid 
apparent discontinuities in redshift not readily seen in the observed data.  While beyond the scope 
of this paper, a complete dynamic model might be needed before more meaningful 
comparisons can be made.  We intend to address this issue in a forthcoming paper, including 
a more rigorous statistical comparison between the predictions of our model and the 
available empirical data, without making any a priori assumptions regarding the particle mass 
function.  

Finally, our proposed mechanism for the observed distance-redshift relation also naturally reproduces other 
observations on large-scales.  For example, at least qualitatively, this mechanism should also produce 
gravitational lensing, and could be empirically-tested via gravitational 
lensing experiments.  Naively, gaps in the particle mass function could translate into discontinuities 
or sharp truncations in the observed enclosed mass as a function of distance from the centre of 
mass of the lensing mass distribution.  The Cosmic Microwave Background can also be qualitatively 
explained within the context of our model.  CMB photons have been traveling at the speed of light since 
the very early Universe  (in $\Lambda$CDM).  Hence, those CMB photons detected at Earth originated 
from the greatest possible distances, and hence the deepest possible potentials (in our model).  Consequently, 
they should be the most redshifted photons in the Universe.  In other words, within the context of our model, CMB 
photons probe the very bottom end of, or minimum particle mass in, the particle mass function.  Hypothetically, 
the observed fluctuations in the energies of CMB photons could constrain the initial 
spatial distribution of the lowest mass particles in the particle mass function.  

\subsubsection{Galactic rotation curves and dark matter} \label{DM}

Interestingly, a potential connection can also be made to dark matter particles via our model.  This could be 
the case if the wavelengths of any particles in our toy model are comparable to or smaller than 
typical galactic scales.  

For instance, consider observed extragalactic rotation curves at large galactocentric radii, which 
tend to be flat as a function of galactocentric 
distance r, attributed to the presence of unseen dark matter particles.  That is, to first order:
\begin{equation}
\label{eqn:vc}
v_{\rm c}^2 = \frac{M(r)}{r} = constant,
\end{equation}
where v$_{\rm c}$ is the circular velocity and M(r) is the enclosed mass at galactocentric radius r.  
Equation~\ref{eqn:vc} constrains the functional form of the particle mass function at large 
galactocentric radii,\footnote{Note that this distance scale should apply to the heaviest particles in, or the "top" 
end of, our assumed particle mass function.} similar to the 
observed distance-redshift relation in Section~\ref{observed}.  To see 
this, we calculate the total mass enclosed within a radius r:
\begin{equation}
\label{eqn:mass2}
\frac{M(r)}{r} \sim \epsilon_{\rm i}\lambda_{\rm i}^2 \sim m_{\rm i}^{1-\alpha}\lambda_{\rm, i}^2  \sim \lambda_{\rm i}^{(\alpha+2\gamma+1)/(\gamma+1)} \sim r^{(\alpha+2\gamma+1)/(\gamma+1)}
\end{equation}
where the second equality holds since we are considering a specified volume.  
In order to reproduce Equation~\ref{eqn:vc}, we thus require ($\alpha$+2$\gamma$+1) $=$ 0 in Equation~\ref{eqn:mass2}, or:
\begin{equation}
\label{eqn:const1}
\alpha + 2\gamma = -1
\end{equation}

The above example illustrates that extragalactic rotation curves could offer an additional pathway toward 
constraining the precise functional form of the particle mass function in our model, provided some particles have 
wavelengths smaller than typical galactic scales.  For example, if a large discontinuity or gap in the mass function is present, 
this could manifest itself observationally if the circular velocity begins to (temporarily) drop off with galactocentric 
distance as 1/r (not including the baryonic mass), instead of v$_{\rm c} =$ constant.  This is because, over some small range in 
r $\sim$ $\lambda_{\rm i}$, (and $\lambda_{\rm i} \ll \lambda_{\rm i+1}$), the mass interior to r is constant with increasing r.  
This 1/r decrease should continue until r $\ge \lambda_{\rm i+1}$, at which point a sharp increase in v$_{\rm c}$ 
could be observed (ignoring the aforementioned oscillating perturbations in Section~\ref{friedmann}).  \textit{We emphasize 
that this proposed observational effect is an artifact of our simple model, and should be confirmed in future studies using 
more sophisticated dynamical modeling.}  We intend to explore in more detail a possible connection between 
the matter distribution presented in this paper and dark matter particles in a future paper.

\subsubsection{Wide binary stars in the Galactic field} \label{wide_binaries}

Another possible test for the extension of our model into the regime of galactic potentials can 
perhaps be constructed from observational monitoring of wide binary stars 
in the Galactic field.  The test works as follows.  If dark matter consists of very long-wavelength 
($\lambda \gtrsim$ 1 pc) particles then, within the context of our simple model, no dark matter should contribute 
to the gravitational potential within the binary's orbit.  If, on the other hand, dark matter consists of particles 
with small wavelengths, then dark matter could exist within the orbits of very wide binaries.  If the mass in dark matter 
is significant, the binary orbit will not be closed and should exhibit deviations from Kepler's Law.  Assuming a continuous 
density of DM particles distributed throughout the Galaxy according to a Navarro-Frenk-White (NFW) profile, 
the total mass of dark matter particles within the orbits of wide low-mass binaries can become comparable 
to or even exceed the total binary mass at certain Galactocentric radii.  This effect could therefore be 
significant at these locations within the Galactic potential.  Additionally, the presence of DM within the 
binary orbit changes the zero-point of the binary orbital energy, such that binaries with separations 
larger than some critical semi-major axis should be disrupted.  In our simple model, however, this is 
only the case provided $\lambda <$ a, where a is the orbital separation of the binary.

\begin{figure}
\begin{center}
\resizebox{!}{95mm}{\includegraphics{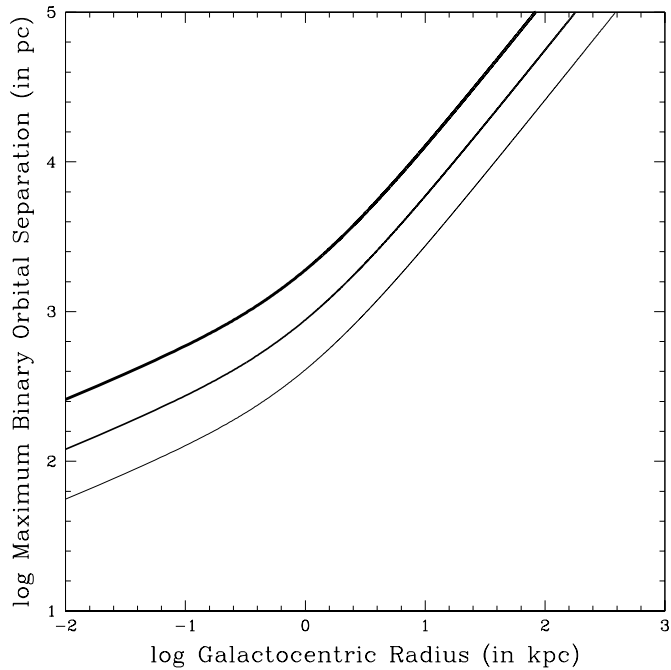}}
\caption{The maximum orbital separation possible for wide binaries in the Galactic field.  We 
assume an NFW density profile for the Milky Way dark matter potential.  We assume equal-mass 
binaries, and show our results for three different binary masses, namely 0.2, 2.0 and 20.0 M$_{\odot}$.  
The line thickness is proportional to the total binary mass (i.e., the thickest line corresponds to the 
most massive binary).  
\label{fig:fig7}}
\end{center}
\end{figure}

We quantify this effect in Figure~\ref{fig:fig7}, which shows the maximum binary orbital separation possible 
for different total binary masses, as a function of Galactocentric radius.  We assume a uniform DM 
density for this calculation.  Specifically, we adopt an NFW profile for the DM 
component of the Galaxy, using the fit parameters in Table 1 of \cite{nesti13}.  As is clear, only in the 
outer reaches of the Galactic halo, where the stellar density is much lower than the local DM density, 
could this effect become significant.  A more viable test might therefore be \textit{galaxy} pairs in 
the outer reaches of galaxy clusters.  We intend to explore this idea further in future work.


\subsection{Caveats and Future Work} \label{future} 

As already discussed, we intend to explore in more detail in a future paper the empirical 
constraints discussed in the preceding sections, which are relevant to large-scale astrophysical 
observations.  \textit{We emphasize that any potentially viable model to describe the large-scale structure of the 
Universe along with its time evolution must satisfy the available wealth of empirical constraints already in hand, while 
ideally also making predictions for future data.}  In this regard, we have only scratched the surface in 
this paper.  However, our model also draws attention to a number of interesting 
issues that could bear important insight for models of quantum gravity.  For example, 
our assumption regarding the nature of Gauss' Law is critical to our model, and 
could potentially be tested in the laboratory.  The assumption that 
gravity can mediate the overlap of wave packets in space and time during gravitational collapse 
is also central to our 
toy model, but remains a subject of active research \cite{das15}.  More generally, 
it is unclear how such long-wavelength particles should interact at all, either 
gravitationally or otherwise.  One of our goals with the toy model presented in this paper is 
to help guide future studies toward key topics that, once better understood, could 
have important implications for future astrophysical 
observations on large spatial scales.  Depending on the validity of our assumptions, the 
model presented in this paper could serve in future studies as a benchmark for extending 
the Standard Model of Particle Physics to very low energy scales.

\section{Summary} \label{summary}

In this paper, we present a novel mechanism to account for the observed distance-redshift relation.  
This is done by presenting a toy model for the large-scale matter distribution in a static 
Universe.  Our model relies on a few key assumptions, including a mass spectrum 
dN$_{\rm i}$/dm$_{\rm i}$ $=$ $\beta$m$^{-\alpha}$ (where $\alpha$ and $\beta$ are 
both positive constants) for low-mass particles with m$_{\rm i}$ $\ll$ M$_{\rm P}$, where M$_{\rm P}$ 
is the Planck mass, and a particle mass-wavelength relation of the form 
$\lambda_{\rm i} =$ $\hbar$/$\delta_{\rm i}$m$_{\rm i}$c, 
where $\delta_{\rm i} =$ $\eta$m$_{\rm i}^{\gamma}$ and $\eta$ and $\gamma$ are both 
constants.   Our model mainly concerns particles with masses far below those in the 
Standard Model of Particle Physics.  For such low-mass particles, we assume that locality can only 
be defined on very large spatial scales, comparable to or exceeding the particle wavelengths.  

We use our model to derive the cosmological redshift characteristic of the Standard 
Model of Cosmology (i.e., $\Lambda$CDM), which becomes a gravitational redshift 
in our toy model.  We then go on to derive Hubble's Law, and show that, within the context of our 
model assumptions, this constrains the particle mass spectrum such that 
$\alpha$ + $\gamma$ $=$ 0 for a linear distance-redshift relation.  
We further compare the results of our model to empirical data and show that, in order to 
reproduce the observed sub-linear form of the distance-redshift relation, our model 
requires $\alpha$ + $\gamma$ $<$ 0.  Taken at face value, the observed data 
also suggest that the particle mass function is relatively continuous, with the maximum 
gap or bin size satisfying m$_{\rm i+1}$/m$_{\rm i}$ $<$ 10$^2$ for successive 
particle masses, for all i (and assuming $\gamma =$ -0.5).  

We further place our toy model in 
the context of the Friedmann Universe, in order to better understand the expected characteristic 
behaviour of a more dynamic version of our model.  Given the overwhelming evidence supporting 
an expanding Universe, we then address possible modifications to our base (static) model that would be 
required to account for the available empirical constraints, including the addition of some initial 
expansion.   Finally, we consider potentially observable distinctions between the cosmological 
redshift and our proposed mechanism to account for the observed distance-redshift relation.  

In conclusion, the mechanism presented here for the observed distance-redshift relation has the 
potential to unify into a single mechanism the source of the 
observable properties of the Universe on large spatial scales, presently attributed to a 
combination of dark matter and dark energy, while also potentially 
offering several unique observational signatures relative to the current Standard Model of 
Cosmology.  

\section*{Acknowledgments}
  
NL would like to thank Solomon Endlich, Achim Kempf, Cliff Burgess, Nick Stone, Leo van Nierop, 
Lauranne Fauvet, Alison Sills, Torsten B\"oker, Dennis Duffin, Jeremiah Ostriker, Mordecai-Mark Mac Low, 
Taeho Ryu and Rosalba Perna for useful discussions and feedback.  O.G. is supported by an NSF Astronomy 
and Astrophysics Fellowship under award AST-1602595.

%


\section*{References} \label{refs}

\begin{thebibliography}{99}

\bibitem{ade15} Ade P. A. R., Aghanim M., Arnaud M., Ashdown J., 
Aumont C., Baccigalupi A. J., Banday R. B., Barreiro J. G., et al. arxiv.org/abs/1502.01589 (2015) 
\bibitem{amanullah10} Amanullah R., Lidman C., 
Rubin D., Aldering G., Astier P., Barbary K., Burns M. S., Conley A., et al., ApJ {\bf 716}, 712 (2010)
\bibitem{burgess04} Burgess C. P. Living Reviews in Relativity {\bf 7}, 5 (2004)
\bibitem{coleman92} Coleman S., Preskill J., Wilczek F. Nuclear Physics B {\bf 378}, 175 (1992)
\bibitem{das15} Das S. Physical Review D {\bf 89}, 8 (2015)
\bibitem{donoghue94} Donoghue J. F. Physical Review D {\bf 50}, 3874 (1994)
\bibitem{donoghue94b} Donoghue J. F. Physical Review Letters {\bf 72}, 2996 (1994)
\bibitem{frieman08} Frieman J. A., Turner M. S., Huterer D. ARA\&A, {\bf 46}, 385 (2008)
\bibitem{hawking74} Hawking S. W. Nature {\bf 248}, 30 (1974)
\bibitem{hawking76} Hawking S. W. Physical Review D {\bf 14}, 2460 (1976)
\bibitem{hinshaw13} Hinshaw G., Larson D., Komatsu E., 
Spergel D. N., Bennett C. L., Dunkley J., Nolta M. R., Halpern M., et al. ApJS {\bf 208}, 19 (2013)
\bibitem{thooft85} t'Hooft G. Nuclear Physics B {\bf 256}, 727 (1985)
\bibitem{hubble29} Hubble E. Proceedings of the National Academy of Sciences of the United States of America {\bf 15}, 168 (1929)
\bibitem{hui16} Hui L., Ostriker J. P., Tremaine S., Witten E. 2016, Physical Review D, submitted (2016; arxiv.org/abs/1610.08297)
\bibitem{landau75} Landau L. D., Lifshitz E. M. 1975, The Classical Theory of Fields (Oxford: Pergamon Press)
\bibitem{lubin01} Lubin L. M., Sandage A. AJ {\bf 122}, 1084 (2001)
\bibitem{marsh15} Marsh D. J. E. arXiv:1510.07633 (2015) 
\bibitem{nesti13} Nesti F., Salucci P. Journal of Cosmology and Astroparticle Physics {\bf 07}, 016 (2013)
\bibitem{oppenheimer39} Oppenheimer J. R., Snyder H. Physical Review {\bf 56}, 455 (1939)
\bibitem{penrose65} Penrose R. Physical Review Letters {\bf 14}, 57 (1965)
\bibitem{penrose69} Penrose R. Rivista del Nuovo Cimento {\bf 1}, 252 (1969)
\bibitem{penzias65} Penzias A. A., Wilson R. W. ApJ {\bf 142}, 419 (1965)
\bibitem{riess04} Riess A. G., Strolger L.-G., Tonry J., Casertano S., Ferguson H. C., 
Mobasher B., Challis P., Filippenko A. V., et al. ApJ {\bf 607}, 665 (2004)
\bibitem{riess09} Riess A. G., Macri L., Li W., Lampeitl H., Casertano S., 
Ferguson H. C., Filippenko A. V., Jha S. W. ApJS {\bf 183}, 109 (2009)
\bibitem{weinberg08} Weinberg S. 2008, Cosmology (Oxford: Oxford University Press)

\end{thebibliography}
\end{document}